\documentclass[12pt]{article}
\usepackage[utf8]{inputenc}
\usepackage{amssymb}
\usepackage{amsfonts}
\usepackage{amsmath}
\usepackage{cite}
\usepackage{comment}
\usepackage{graphicx}
\usepackage{multicol}
\usepackage{multirow}
\usepackage{setspace}
\usepackage{lineno}
\usepackage{makecell}
\usepackage{natbib}
\usepackage{caption}
\usepackage{url}
\usepackage{color}
\usepackage{layouts}
\setcitestyle{authoryear}
\usepackage[vmargin=1in,hmargin=1in]{geometry}

\newgeometry{hmargin=1in}

\title{\large \textbf{Smooth Concordance Metrics for Survival Models}}
\author{\large Nicholas Hartman and Grace Richards}

\date{\normalsize Department of Biostatistics, University of Michigan, Ann Arbor, MI, U.S.A.}

\begin{document}

\maketitle

\begin{abstract}
Concordance indices are widely popular metrics for assessing the ability of predictive survival models to discriminate underlying risk levels. However, these statistics have also been criticized for using only the rank orderings of the model’s predicted risk scores and being insensitive to important model features, such as the addition of strong predictor variables into the model. In this paper, we address these limitations by developing smooth concordance metrics that model the underlying risk discrimination probabilities as continuous functions of the predicted risk score differences, where the shapes of these functions are estimated from the observed data. As a result, these smooth concordance metrics assess model performance across the entire range of possible risk score differences, allowing one to identify specific scenarios where the candidate model performs especially well or better than other models. Simulations show that the proposed smooth concordance metrics provide more detailed information about risk discrimination performance and are much more sensitive to the addition of meaningful predictors. We apply these methods to compare predictive survival models for cancer recurrence.  

\vspace{14pt}

\end{abstract}

\clearpage

\singlespacing

\section{Introduction}
\label{sec:intro}

Predictive survival models have proven to be useful across many applications where it is necessary to allocate resources to high-risk individuals or make decisions based on predicted survival probabilities \citep{Steyerberg}. For example, survival models have been integrated in clinical contexts to help make accurate prognoses and provide individualized and preemptive healthcare. To ensure that these predictions are effective, model developers must first validate models with various performance metrics related to calibration (i.e., the accuracy of predicted probabilities) or discrimination (i.e., the ability to correctly distinguish higher versus lower risk individuals) or both \citep{DeMaris}. In this paper, we focus on one of the most popular metrics of discrimination performance for survival models, known as the concordance index (C-Index), which describes the probability of correctly ordering patients’ survival times based on the ordering of the predicted risk scores \citep{Harrell,Harrell1982}.

While the C-Index is used widely in practice, it has also faced substantial criticism for its limitations in describing key aspects of risk discrimination performance for survival outcomes \citep{Cook,Vickers2,Vickers,Halligan,Hartman}. Specifically, because the C-Index focuses solely on the rank ordering of the predicted risk scores as opposed to their actual values, it is known to be highly insensitive to the addition of clinically important predictors with large effect sizes. In addition, it has been shown that the C-Index is often dominated by comparisons of subjects with very similar risk profiles and therefore fails to capture useful discrimination performance among more clinically meaningful underlying risk differences. Thus, it is common for the C-Index to severely understate the utility of survival models in relevant scenarios for clinical decision-making \citep{Hartman}. 

To address these limitations, we propose a novel smooth concordance metric that allows for assessments of survival model discrimination performance across the entire spectrum of risk score differences. By leveraging this smooth concordance metric, one can easily characterize various levels of granularity in risk discrimination achieved by the model and identify thresholds of risk score differences where the model is most useful at distinguishing future survival experiences. We also show that the proposed smooth concordance metric can be estimated based on a weighted generalized estimating equation applied after an initial data pre-processing step, allowing for convenient implementation and straightforward derivations of the asymptotic properties based on familiar methods. Simulations show that the proposed metric is much more sensitive to the addition of clinically meaningful predictors and provides substantially more information on model performance compared to the traditional C-Index. We demonstrate these differences in metrics through a real study of breast cancer recurrence. 

\section{Notation}

Let $T_i$ denote the underlying failure time and $D_i$ denote the underlying right-censoring time for the $i$th individual in the sample. We observe the follow-up time $X_i=\min(T_i,D_i)$ and event indicator $\delta_i=I(T_i < D_i)$, where the indicator function $I(\cdot)$ equals one if the condition is true and zero otherwise. Furthermore, assume that a risk score $R_i$ is measured for each individual based on a vector of predictors $\boldsymbol{Z}_i$. Without loss of generality, we define $R_i=\boldsymbol{Z}_i^{\top}\boldsymbol{\beta}$, where $\boldsymbol{\beta}$ is a vector of model coefficients (e.g., from a Cox proportional hazards model), but any arbitrary risk score is compatible with the methods discussed in this paper. For a sample of $n$ subjects, we collect data $\{X_i, \delta_i, R_i\}, i=1,\dots,n$.

\section{Background}

\subsection{A Review of the Traditional C-Index}

We briefly review the traditional C-Index and its limitations, which have been discussed extensively elsewhere \citep{Pencina,Halligan,Hartman}. The C-Index is an estimator for the probability that an individual with an earlier underlying failure time would have a higher predicted risk score from the model compared to another individual with a later underlying failure time. This quantity is known as the concordance probability, and it serves as the population parameter of interest. Mathematically, it is written as \begin{equation}
    P(R_i > R_j | T_i < T_j).
    \label{eq:target}
\end{equation}

\noindent A concordance probability of 0.5 corresponds to a random ordering of patient risk levels and represents the minimum performance. \citet{Harrell1982} proposed the original C-Index estimator as the fraction of comparable pairs (pairs in which the rank ordering of underlying failure times is discernible) that have correctly ordered or “concordant” risk scores: \begin{equation}
    \widehat{C}_H=\frac{\sum_{i \ne j} I(X_i <X_j, \delta_i=1,R_i > R_j)}{\sum_{i \ne j} I(X_i <X_j, \delta_i=1)}.
\end{equation}

Many extensions of Harrell’s C-Index have been proposed, such as Uno’s C-Index which incorporates inverse probability of censoring weights to correct for bias due to right-censoring \citep{Uno}. One property of the C-Index is that it defines concordance based on the ordering of the risk scores and uses indicator functions to check the sign of the differences in risk scores within each comparable pair. In this way, it dichotomizes a continuous random variable into a binary one, which uses less information and implicitly assumes that only the ordering of the risk scores is relevant for risk discrimination purposes. Previous work has shown that this reduction of information is responsible for the C-Index being highly insensitive to the addition of useful predictors and overemphasizing sets of comparable pairs with clinically meaningless differences in risk profiles \citep{Cook,Hartman}. 

\subsection{Direct C-Index Optimization}
\label{sec:direct}

\begin{figure}[h!]
    \centering
    \includegraphics[width=\textwidth]{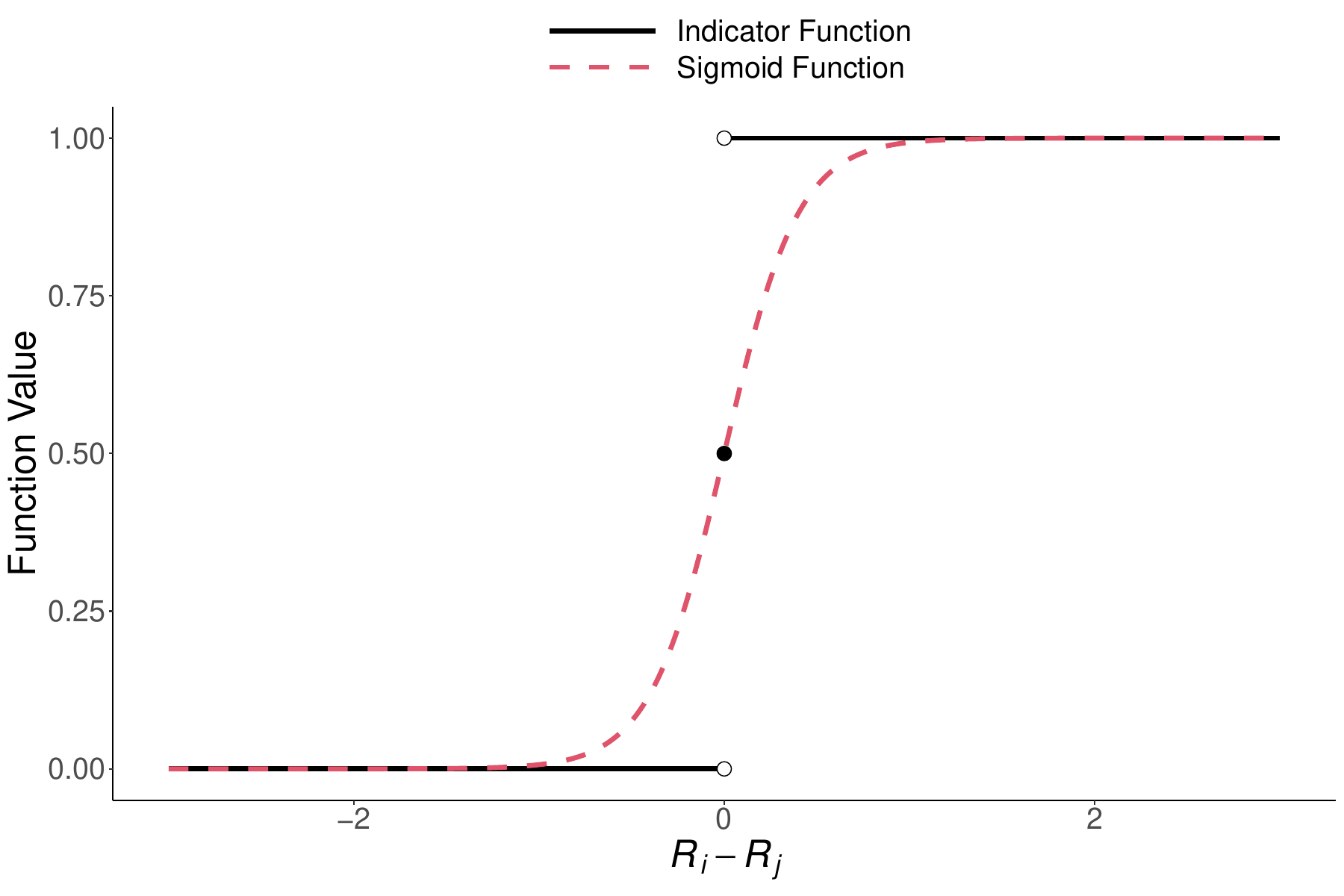}
    \caption{An example sigmoid function approximation to the indicator function from the traditional C-Index formula, as used in direct C-Index optimization. Ties in the risk scores $R_i$ and $R_j$ are typically scored with 0.5 in the traditional C-Index formula.}
    \label{fig:sigmoid}
\end{figure}

The concept of a smooth concordance metric was first considered for the purpose of fitting survival models that directly optimize the traditional C-Index \citep{Chen,Mayr}. That is, instead of finding the $\boldsymbol{\beta}$ values that maximize the partial likelihood of a Cox proportional hazards model (or the likelihood of any other survival model), it has been suggested that one could find the $\boldsymbol{\beta}$ values that produce the largest C-Index value. However, given that the traditional C-Index is based on a series of indicator functions, it is non-differentiable with respect to $\boldsymbol{\beta}$. Thus, to define a more well-behaved objective function for direct C-Index optimization, \citet{Chen} proposed the following smooth approximation of the C-Index based on sigmoid functions: \begin{equation}
\widehat{C}_S(\boldsymbol{\beta})=\frac{\sum_{i \ne j} I(X_i < X_j, \delta_i=1)\bigg\{\frac{\exp(\nu(\boldsymbol{Z}_i^{\top}\boldsymbol{\beta}-\boldsymbol{Z}_j^{\top}\boldsymbol{\beta})}{1+\exp(\nu(\boldsymbol{Z}_i^{\top}\boldsymbol{\beta}-\boldsymbol{Z}_j^{\top}\boldsymbol{\beta})}\bigg\}}{\sum_{i \ne j} I(X_i <X_j, \delta_i=1)}.
    \label{eq:smooth_optim}
\end{equation}

\noindent where $\nu$ is a tuning parameter that is prespecified such that the sigmoid function closely resembles the original indicator function in the traditional C-Index (Figure \ref{fig:sigmoid}).

\subsection{Delta Concordance Probability}
\label{sec:dcp}

Motivated by similar goals as in this paper, \citet{Devlin} explored a modified definition of the concordance probability, which they referred to as the ``Delta Concordance Probability (dCP),'' to emphasize clinically meaningful comparisons in the concordance calculations:
\begin{equation}
    P(R_i > R_j | T_i + \Delta < T_j),
    \label{eq:delta_target}
\end{equation}

\noindent where $\Delta$ is some user-defined difference in underlying failure times that is deemed clinically meaningful. They then proposed a consistent estimator for (\ref{eq:delta_target}) under various model assumptions, and explored the reporting of this estimator across a range of $\Delta$ values, which showed improved discrimination performance for pairs with more clinically meaningful differences in underlying failure times. 

While \citet{Devlin}'s estimator is useful for addressing many of the practical issues with the C-Index described in Section \ref{sec:intro}, it has two key limitations that may impact its applicability. First, the proposed concordance estimator depends on model-based estimates of the conditional survival function, and therefore requires a correctly-specified survival model that can generate accurate survival probabilities. However, in practice, it is usually the case that candidate risk prediction models are underspecified, making it difficult to achieve this required condition for valid model evaluations. Second, the dCP conditions on clinically meaningful differences in underlying failure times and assesses the ordering of predicted risk scores. However, when risk prediction models are used in real clinical settings, the underlying failure times are unknown. Thus, for policymakers who make decisions about healthcare resource allocation based on predictive models, it may be more informative to consider the probability that one patient will fail before another, conditional on an observed difference in predicted risk scores. In this way, the concordance metric would give insight into which magnitudes of risk score difference are meaningful enough to have confidence in the future ordering of the underlying failure times. 

\section{The Proposed Smooth Concordance Metric}

\subsection{Methodology}

We aim to build upon the ideas described in Sections \ref{sec:direct} and \ref{sec:dcp} to address the practical issues of the C-Index and the limitations of \citet{Devlin}'s concordance estimator. The smooth concordance metrics as described in Section \ref{sec:direct} have thus far only been considered specifically for the purpose of finding covariate effects $\boldsymbol{\beta}$ that optimize the traditional concordance probability. However, in this paper, we are motivated by an entirely different purpose for smooth concordance metrics that requires new methodological considerations. In our setting of interest, we assume that a risk score (e.g., $R_i=\boldsymbol{Z}_i^{\top}\widehat{\boldsymbol{\beta}}$) has already been estimated based on a fitted survival model, and the goal is then to estimate the shape of the smooth sigmoid function that best describes the predictive performance of this risk score. Thus, instead of forcing the performance metric to only depend on the rank ordering of the risk scores as in the traditional C-Index, we use the observed data to flexibly characterize  model performance as a continuous function over the entire range of differences in risk scores. In addition, here we treat the $\nu$ parameter and the sigmoid function itself as the main performance metrics of interest to estimate from the data, as opposed to using $\nu$ solely as an intermediate tuning parameter that is specified to ultimately calculate a metric similar to the traditional C-Index as in \citet{Chen}. 

To define our smooth concordance metric, we first re-express the concordance probability as $P(T_i < T_j | R_i > R_j)$ \citep{GH}. Then, we make this quantity more flexible by conditioning instead on the difference in risk score values and modeling it through a sigmoid function with parameter $\nu$: \begin{equation} P(T_i < T_j | R_i-R_j)=\frac{\exp(\nu h(R_i-R_j))}{1+\exp(\nu h(R_i-R_j))}, \label{eq:proposed} \end{equation}

\noindent where $h(\cdot)$ is any user-defined monotonic function. For simplicity, we introduce our methods with $h(\cdot)$ defined as the identity function such that $h(R_i-R_j)=R_i-R_j$. For the special case of a perfectly specified Cox proportional hazards model with $R_i=\boldsymbol{Z}_i^{\top}\boldsymbol{\beta}$, (\ref{eq:proposed}) holds with $h(R_i-R_j)=R_i-R_j$ and $\nu=1$, which provides additional motivation for the use of the sigmoid function \citep{DeNeve}. However, even for a risk score derived from an unknown or misspecified survival model, we argue that the assumption in (\ref{eq:proposed}) is relatively mild compared to estimators that require correct specification of the conditional survival functions, such as \citet{Devlin}. As discussed in Section \ref{sec:dcp}, our proposed estimator conditions on differences in the predicted risk scores, as opposed to differences in underlying failure times, which may help identify specific patient pairs at the time of prediction for which the model is expected to perform better. 

It follows that the quantity $\exp(\nu)$ is the odds ratio describing the multiplicative change in odds of patient $i$ failing before patient $j$ for every one-unit increase in the risk score difference $R_i-R_j$. Larger values of $\exp(\nu)$ indicate that the model's ability to discriminate underlying risk (and correctly order event times) increases more rapidly with the risk score differences $R_i-R_j$, suggesting better model performance. In addition, plots of the sigmoid function itself provide information on how the model's concordance rates change with the differences in predicted risk scores. From this information, one can then identify the amount of difference in predicted risk scores that is needed to have sufficient confidence in the model's ability to distinguish underlying risk. 

\begin{figure}[h!]
    \centering
    \includegraphics[width=\textwidth]{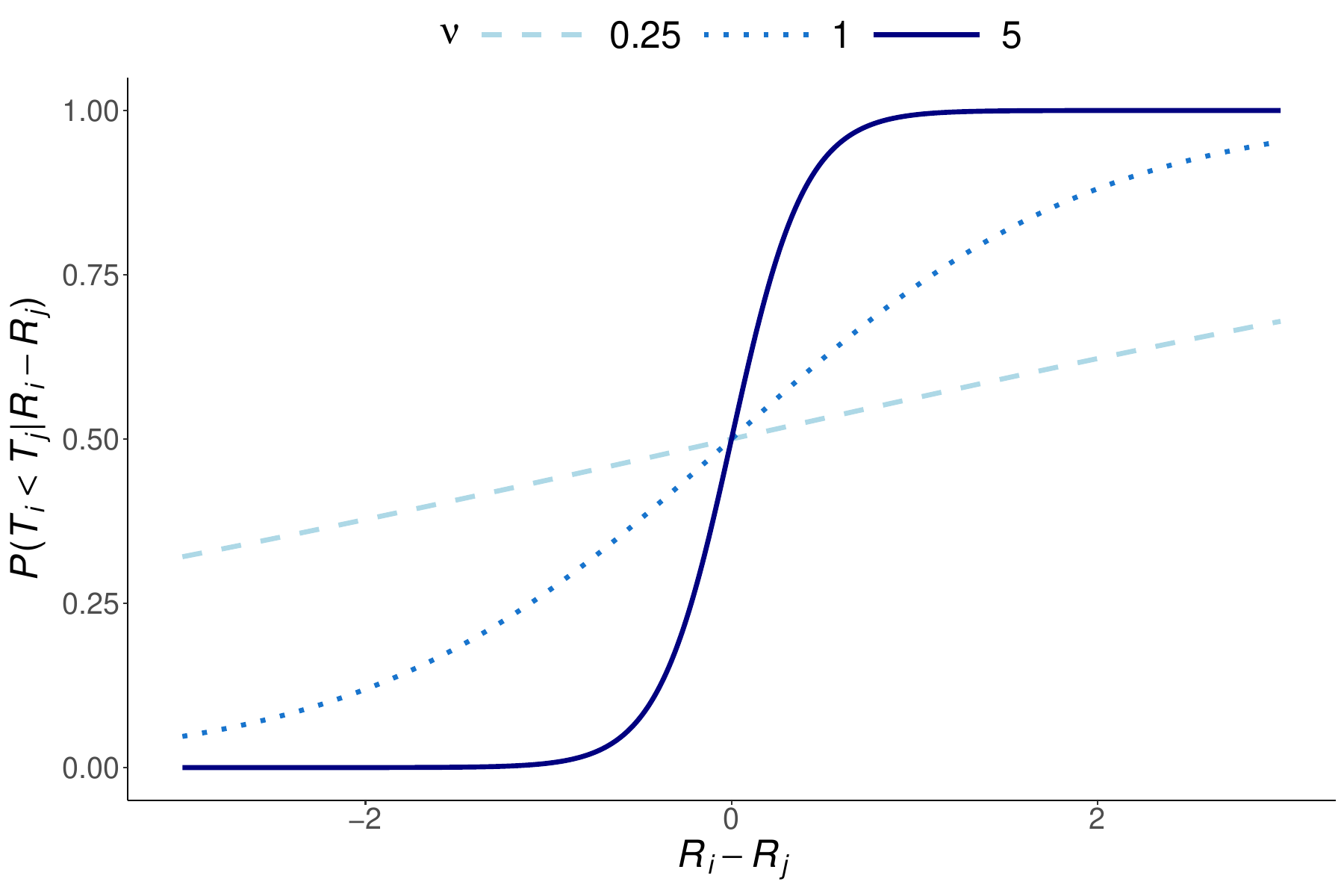}
    \caption{A visualization of the proposed smooth concordance function that relates the probability of ordering the failure times $T_i$ and $T_j$ with the difference in risk score $R_i-R_j$. The parameter $\nu$ controls the shape of the function, and is estimated from the data. Larger values of $\nu$ suggest that the risk scores are better able to discriminate underlying risk.}
    \label{fig:sigmoid2}
\end{figure}

\subsection{Estimation}
\label{sec:estim}

It is easy to recognize that the proposed smooth concordance relationship in (\ref{eq:proposed}) aligns with a semi-parametric transformation model specification known as a Probability Index Model \citep{Dodd,Thas,DeNeve}, with the risk score difference $R_i-R_j$ as the sole predictor. Therefore, the estimation of the main parameter of interest $\nu$ follows easily from existing generalized estimating equation (GEE) approaches that account for both the right-censoring and correlation between pairwise outcomes \citep{Dodd,Cheng}. Similar to Uno's C-Index \citep{Uno}, the estimation strategy proceeds by only using comparable pairs of subjects that have a known ordering of failure times (i.e., $X_i < X_j$ where $\delta_i$=1 or $X_i > X_j$ where $\delta_j$=1) and leveraging inverse probability of censoring weights (IPCW) in the estimating equation.

First, define $\boldsymbol{W}_i$ as a diagonal weight matrix of dimension $m_i \times m_i$, where $m_i$ is the number of comparable pairs involving the $i$th subject. The $k$th diagonal element of $\boldsymbol{W}_i$ is $G(\min\{X_i,X_{j[k]}\} | \boldsymbol{Z}_i)^{-1}G(\min\{X_i,X_{j[k]}\} | \boldsymbol{Z}_{j[k]})^{-1}$, where $G(t | \boldsymbol{Z})=P(D > t | \boldsymbol{Z})$ describes the conditional censoring distribution and the $j[k]$ subscript denotes the index of the subject that forms the $k$th comparable pair involving the $i$th subject. The function $G(t | \boldsymbol{Z})$ can be consistently estimated from a correctly specified survival model with follow-up times $X_i$ and censoring indicator $1-\delta_i$. Furthermore, let $\boldsymbol{Y}_i$,  $\boldsymbol{\mu}_i$, and $\boldsymbol{Q}_i$ be $m_i$-dimensional column vectors having $k$th elements of $I(X_i < X_{j[k]})$, $\exp(\nu h(R_i-R_{j[k]}))/\{1+\exp(\nu h(R_i-R_{j[k]}))\}$, and $\partial \mu_{ik} / \partial \nu$ respectively. Finally, denote $\boldsymbol{\Sigma}_i$ as a working covariance matrix with dimension $m_i \times m_i$. 

Under conditionally-independent right-censoring (i.e., $T_i$ and $D_i$ are independent conditional on $\boldsymbol{Z}_i$), the weighted estimating equation of interest, as defined in \citet{Cheng}, is \begin{equation}
U(\nu)=\sum_{i=1}^{n} \boldsymbol{Q}_i^{\top}\boldsymbol{\Sigma}_i^{-1}\boldsymbol{W}_i\left(\boldsymbol{Y}_i - \boldsymbol{\mu}_i \right)=0.
\label{eq:estim}
\end{equation}

\noindent It is known from established GEE and IPCW theory that the $\widehat{\nu}$ which solves the weighted estimating equation described above in (\ref{eq:estim}) is a consistent estimator for $\nu$ in the presence of right-censoring with a correctly specified censoring model \citep{Preisser,Blanche}. In summary, the estimation of $\nu$ can be implemented through the following steps in practice:

\begin{enumerate}
 \item Using the observed data $\{X_i,\delta_i,R_i\}$, estimate the function $G(t|\boldsymbol{Z})$ based on a survival model with follow-up times $X_i$ and censoring indicator $1-\delta_i$.
 \item Sort the data $\{X_i,\delta_i,R_i\}$ by the risk scores $R_i$.
 \item From the sorted data, enumerate all pairs of subjects and record the difference in risk scores $R_i-R_j$ for each pair. 
 \item Subset the dataset of pairs to only include comparable pairs with $X_i < X_j$ and $\delta_i=1$ or $X_i > X_j$ and $\delta_j=1$, such that the ordering of $T_i$ and $T_j$ are known. 
 \item Using the subset of comparable pairs, define an outcome variable for each pair as $Y_{ij}=I(X_i < X_j)=I(T_i < T_j)$, to indicate whether subject $i$ failed before subject $j$.
 \item Fit a weighted marginal model with GEE using a logit link function and the dataset of comparable pairs with outcome $Y_{ij}$, predictor $R_i-R_j$, weights $\widehat{G}(\min\{X_i,X_j\} | \boldsymbol{Z}_i)^{-1} \times \widehat{G}(\min\{X_i,X_j\} | \boldsymbol{Z}_j)^{-1}$, and no intercept.
\end{enumerate}

\subsection{Theoretical Properties}
\label{sec:theory}

Given that the proposed estimation strategy can be reduced to a weighted GEE approach, as described in the previous subsection, it follows directly that $\widehat{\nu} \xrightarrow{p} \nu$ and $\sqrt{n}\{\widehat{\nu}-\nu\} \xrightarrow{d} N(0,V)$. In large samples, one may consider approximating the asymptotic variance $V$ based on a robust sandwich estimator that treats the weights as known \citep{Cheng,
Preisser,DeNeve,Blanche}. Alternatively, one may apply a nonparametric bootstrap approach for variance estimation that samples the subjects with replacement and repeats the weight calculation and model fitting for a large number of iterations. While the nonparametric bootstrap approach may better account for any statistical uncertainty in the weights, it has been shown that ignoring this uncertainty typically has very little impact on the standard errors \citep{Azarang}, allowing for a simpler approximation in many cases based directly on the weighted GEE output. 

In addition to the properties described above, we highlight a key mathematical connection between the proposed smooth concordance function and the traditional C-Index. That is, the traditional concordance probability can be written as \begin{equation}
    P(T_i < T_j | R_i > R_j)=\int_0^\infty P(T_i < T_j | R_i-R_j=r_i-r_j) f_{R_i-R_j | R_i > R_j}(r_i-r_j)\,\, d\{r_i-r_j\},
\end{equation}

\noindent where $f_{R_i-R_j | R_i > R_j}(r_i-r_j)$ is the density of the risk score difference, conditional on this difference being positive. Through our smooth concordance function in (\ref{eq:proposed}), we model the conditional probability of ordering the failure times based on the difference in risk scores, $P(T_i < T_j | R_i-R_j)$, and the C-Index is a marginal statistic that averages this conditional probability over the distribution of positive risk score differences. In many applications, this distribution will have the highest density at values that are not clinically meaningful, so the resulting C-Index often fails to reflect important aspects of model performance \citep{Hartman}. By estimating the conditional probability $P(T_i < T_j |R_i-R_j)$ directly and modeling the way in which this probability changes with $R_i-R_j$ (through the $\nu$ parameter), we are able to discern more information about the model's discrimination abilities across various risk difference levels. 

\begin{figure}[h!]
\includegraphics[width=\textwidth]{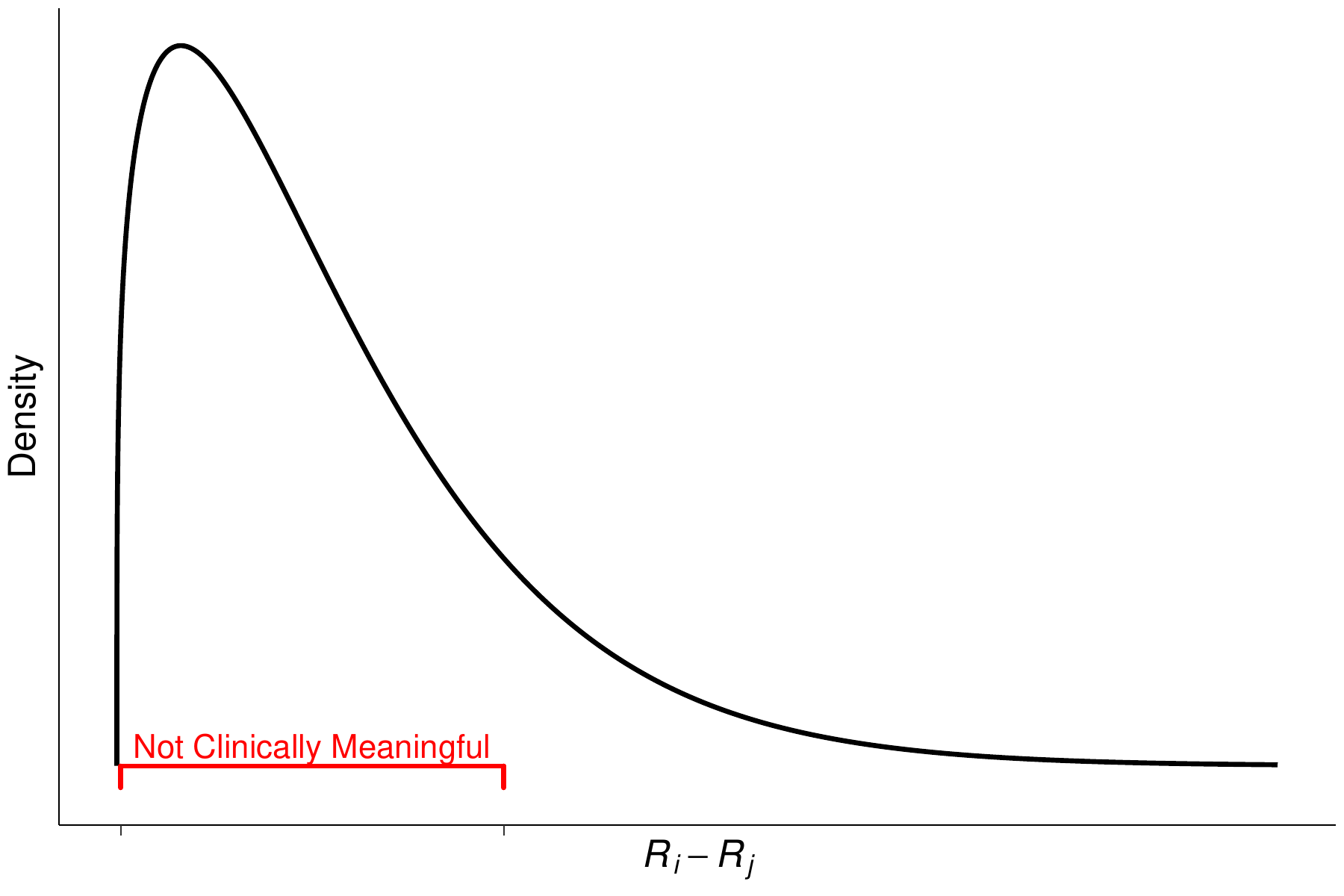}
\caption{An example distribution of pairwise risk score differences $R_i-R_j$, where some of the most commonly observed risk score differences are not clinically meaningful. The traditional C-Index is dominated by the risk score differences that are most commonly observed, whereas the proposed smooth concordance metric describes performance across the entire range of risk score differences.}
\label{fig:R}
\end{figure}

Figure \ref{fig:R} shows an example distribution of risk score differences that helps demonstrate the advantages of the proposed smooth concordance metrics. The distribution of $R_i-R_j$ in this example is highly skewed, which is common in practice especially when the original risk scores have a density function with a tall peak and long tails. Many of the most commonly observed risk differences will not be clinically meaningful, but these differences have the most influence on the traditional C-Index estimate. In contrast, our proposed estimation of the sigmoid function describes the discrimination performance across all values of $R_i-R_j$, and the $\widehat{\nu}$ parameter estimate quantifies the way in which this performance changes with each one-unit increase in $R_i-R_j$.

\section{Simulations}

\subsection{Study 1: Methodological Comparisons}
\label{sec:study1}

To compare the properties of the traditional C-Index and our proposed smooth concordance metric, we conducted a numerical study on simulated survival data with predictors $Z_1$ and $Z_2$. The predictor $Z_1$ was sampled from a standard normal distribution across all settings, and the predictor $Z_2$ was generated from different probability density functions under three separate cases: 

\begin{enumerate}
    \item Case I (Symmetric): $Z_2$ was sampled from a standard normal  distribution, $N(0,1)$
    \item Case II (Skewed): $Z_2$ was sampled from an exponential distribution with rate parameter $\lambda=2.5$
    \item Case III (Outlier Contaminated): 90\% of observations of $Z_2$ were sampled from $N(0,1)$, and 10\% of observations of $Z_2$ were sampled from $N(2, 0.1)$
\end{enumerate}

\noindent In Cases II and III, the distributions of the risk scores are highly skewed with a tall peak and long tail. Therefore, even though $Z_2$ has a meaningful effect size, the risk scores involving this variable will generate many comparable pairs with very similar risk profiles. As a result, we expected that the traditional C-Index would be less sensitive to the addition of $Z_2$ under these cases because it would be overwhelmed by pairs of individuals with very similar risks. 

We chose a study end time of $\tau=30$ and a sample size of $n=250$ for each dataset. In this initial simulation study, the subjects were only right-censored administratively at the end of the study, though we explored other types of right-censoring in later studies. On each iteration of the simulation, we first simulated $Z_1$ and $Z_2$ from the distributions described above. Then, we generated event times, $T$, from a proportional hazards model with a constant baseline hazard of 0.1 and true coefficient values of $\beta_1=0.2$ and $\beta_2=1$ for $Z_1$ and $Z_2$ respectively. We then let $X_i=\min(T_i, \tau)$ and $\delta_i=I(T_i<\tau)$ for each observation, and calculated both the traditional C-index and our proposed smooth concordance metric for risk scores based on two nested Cox regression models: one that included only $Z_1$ as a predictor and another that included both $Z_1$ and $Z_2$ as predictors. The risk scores were defined as the linear predictors $R=\boldsymbol{Z}^{\top}\widehat{\boldsymbol{\beta}}$ from the estimated Cox models. Finally, we computed both metrics for each model and simulated data set, and averaged the estimated values across all 10,000 iterations. This simulation process was carried out separately for each of the three cases described above.

Based on these average concordance metric values across simulations, we also graphed the probability of the first subject in a comparable pair having an earlier underlying failure time (conditional on the difference in predicted risk scores) versus the difference in predicted risk scores, for both the initial model (with only $Z_1$) and the full model (with both $Z_1$ and $Z_2$) across all three cases. This was repeated for both the traditional C-Index and smooth concordance metric, using their respective assumed model forms for $P(T_i < T_j | R_i-R_j)$. That is, Harrell's C-Index corresponds to a piecewise constant function of the risk score differences, defined as: 
\begin{equation}
P(T_i < T_j | R_i-R_j) = 
\begin{cases}
    \widehat{C}_H & R_i > R_j \\
    1-\widehat{C}_H & R_i < R_j
\end{cases},
\end{equation}

\noindent whereas the proposed smooth concordance metric corresponds to a sigmoid function with parameter estimate $\widehat{\nu}$, as described in (\ref{eq:proposed}).

Table \ref{tab:Simulation_Tab1} shows the average C-Index estimate and $\exp(\widehat{\nu})$ value for the underspecified model (with $Z_1$ only) and the full model (with $Z_1$ and $Z_2$) based on each of the three data-generating distributions for $Z_2$. As expected, the traditional C-Index was not very sensitive to the addition of $Z_2$ into the model when this variable was generated under Cases II and III (i.e., when the comparable pairs tended to be dominated by comparisons of very similar risk profiles). Specifically, $\widehat{C}_H$ increased by 35\% under Case I, 11\% under Case II, and 13\% under Case III after adding $Z_2$ into the model. In contrast, the estimated sigmoid function parameter $\exp(\widehat{\nu})$ showed meaningful increases on the odds ratio scale across all three cases. That is, $\exp(\widehat{\nu})$ increased by 252\% under Case I, 48\% under Case II, and 98\% under Case III after adding $Z_2$ into the model. 

\begin{table}[h!]
    \centering
    \caption{Average value of the traditional C-Index estimate ($\widehat{C}_H$) and smooth concordance function parameter on the odds ratio scale ($\exp(\widehat{\nu})$), for models with either one predictor ($Z_1$) or two predictors ($Z_1$ and $Z_2$). Case I: $Z_2$ is simulated from a normal distribution, Case II: $Z_2$ is simulated from a skewed exponential distribution, and Case III: $Z_2$ is simulated from an outlier-contaminated normal distribution. Based on 10,000 simulation iterations.}
    \begin{tabular}{ccccccc}
    \hline
        & \multicolumn{2}{c}{\textbf{Case I}} & \multicolumn{2}{c}{\textbf{Case II}} & \multicolumn{2}{c}{\textbf{Case III}} \\
        \hline
        & $Z_1$ & $Z_1$ and $Z_2$ & $Z_1$ & $Z_1$ and $Z_2$ & $Z_1$ & $Z_1$ and $Z_2$ \\
        \hline
        $\widehat{C}_H$ & 0.54 & 0.73 & 0.55 & 0.61 & 0.55 & 0.62 \\
        $\exp(\widehat{\nu})$ & 1.25 & 4.40 & 1.31 & 1.94 & 1.31 & 2.60 \\
        \hline
    \end{tabular}
    \label{tab:Simulation_Tab1}
\end{table}

\begin{figure}[p]
\centering
\includegraphics[width=0.85\textwidth]{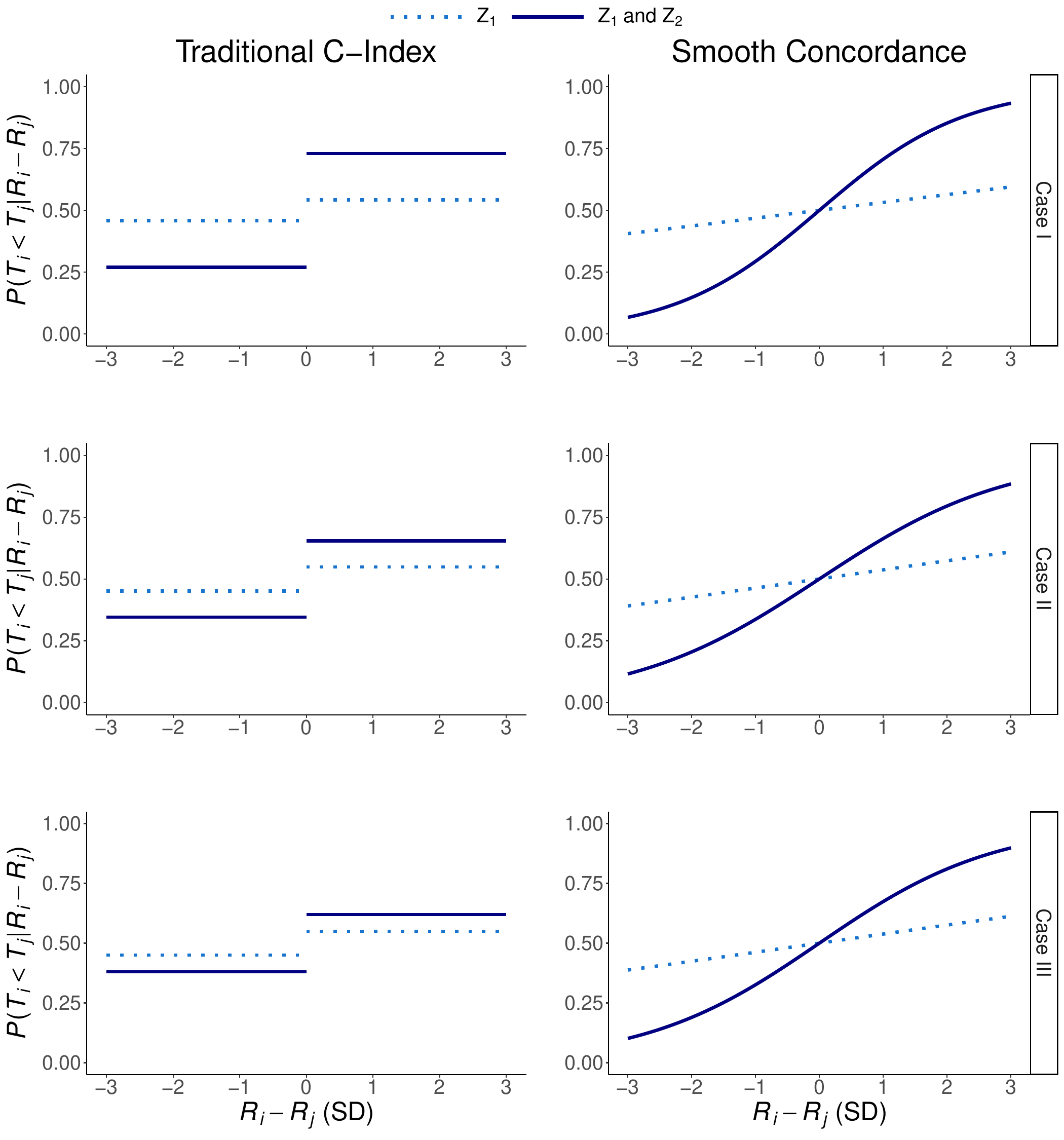}
\caption{Probabilities of subject $i$ having an earlier underlying failure time than subject $j$, based on the difference in risk scores $R_i-R_j$, for models with either one predictor ($Z_1$) or two predictors ($Z_1$ and $Z_2$). The first column  represents the function assumed by the traditional C-Index, whereas the second column represents the sigmoid function under the smooth concordance method. Case I: $Z_2$ is simulated from a normal distribution, Case II: $Z_2$ is simulated from a skewed exponential distribution, and Case III: $Z_2$ is simulated from an outlier-contaminated normal distribution. Based on 10,000 simulation iterations. SD: Standard Deviation.}
\label{fig:Simulation_Fig1}
\end{figure}

Figure \ref{fig:Simulation_Fig1} provides further intuition for the results observed in Table \ref{tab:Simulation_Tab1} by plotting the concordance rates as a function of the risk score differences, based on the functional forms assumed by the traditional C-Index and smooth concordance metric. For Case I, where we added a standard normal predictor $Z_2$ to an initial model including $Z_1$, there were very large increases in concordance rates after the addition of $Z_2$, and this was true for both the traditional C-Index and smooth concordance metric (Figure \ref{fig:Simulation_Fig1}). In Case II, where $Z_2$ followed a highly skewed exponential distribution, the addition of $Z_2$ into the model did not increase the traditional C-Index very much, which was expected because the comparable pairs very frequently involved two subjects with similar risk scores. However, it is evident from the plot of the smooth sigmoid function that this additional predictor greatly improved the model’s risk discrimination ability for subjects with very different profiles. This advantage can be visualized by the large increase in concordance probability from the tails of the sigmoid function (Figure \ref{fig:Simulation_Fig1}). For Case III, where $Z_2$ followed an outlier-contaminated normal distribution with limited variability in non-outlier points, we observed a similar pattern as in Case II. That is, the traditional C-Index failed to capture the predictive value of $Z_2$ because it was dominated by clinically meaningless comparisons between non-outlier points, but the smooth concordance metric clearly showed that the predictor was extremely useful for discriminating earlier versus later failure times among subjects with larger differences in risk profiles (e.g., non-outlier versus outlier). Thus, by plotting the concordance rates over the entire range of risk differences in this way, we are able to consider how the magnitude of the risk score difference is related to the model's discrimination performance.

\subsection{Study 2: Inverse-Probability of Censoring Weights}
\label{sec:study2}

In a second simulation study, we introduced non-administrative right-censoring and confirmed the ability of the IPCW corrections to reduce censoring-related bias in the smooth concordance estimates. For the purpose of this simulation, we compared the average estimates of $\widehat{\nu}$ based on the naive unweighted approach using the right-censored data, the IPCW approach using the right-censored data, and the unweighted approach using the underlying failure times (before non-administrative right-censoring is introduced). The average estimates from this last approach represent the target parameter values if the impact of right-censoring could be eliminated, and ideally the estimates based on the right-censored data would be close to these target values. Thus, we calculated the percent bias for the unweighted and weighted metrics (estimated using the right-censored data) by treating the average metric estimated from the underlying failure times as the true parameter value. 

We simulated the failure time observations in a similar way as in the previous subsection, with just one predictor $Z_1$ sampled from an Exponential distribution with rate parameter $\lambda=1$. Then, we generated the underlying right-censoring variable, $D$, based on a Cox model with constant baseline hazard $\lambda_D$ and coefficient $\beta_D$ for $Z_1$, which we varied to explore different censoring distributions. The observed follow-up times and event indicators were defined as $X_i=\min(T_i, D_i, \tau)$ and $\delta_i=I(T_i < \min(D_i,\tau))$. For the calculations based on the underlying data (before the non-administrative censoring was removed), we re-defined $X_i$ and $\delta_i$ by setting $D_i=\infty$. On each iteration of the simulation, we estimated the $\nu$ parameter for our proposed smooth concordance metric using the estimated risk scores based on $Z_1$ (as outlined in the previous section). 

As expected, the average IPCW estimates based on the right-censored data were closer to the target parameter values than the naive unweighted estimates. In particular, the naive unweighted estimates became substantially more biased as the censoring rate or $\beta_D$ increased, whereas the IPCW method was mostly insensitive to these parameters and remained close to the target values. Thus, these simulations confirm the theory established in this paper and previous works that have shown IPCW methods are capable of reducing bias due to right-censoring if the weights are correctly specified. 

\begin{table}[h!]
    \centering
    \caption{Percent bias in the smooth concordance parameter estimate $\widehat{\nu}$, based on a naive unweighted version and an inverse-probability of censoring weighted (IPCW) version to account for right-censoring. Based on 10,000 simulation iterations.}
    \begin{tabular}{cccc}
    \hline
    $\boldsymbol{\beta_D}$ & \textbf{Censoring Rate} & \textbf{Unweighted} & \textbf{IPCW} \\
    \hline
    1 & 11\% & 0.22 & 0.24 \\
    1 & 42\% & 1.20 & 0.98 \\
    1 & 58\% & 2.39 & 0.49 \\
    1 & 70\% & 4.92 & 0.53 \\
    \hline
    $\boldsymbol{\beta_D}$ & \textbf{Censoring Rate} & \textbf{Unweighted} & \textbf{IPCW} \\
    \hline
    0.5 & 70\% & 1.56 & 0.31 \\
    1.0 & 70\% & 4.22 & 1.42 \\
    1.5 & 70\% & 7.78 & 2.75 \\
    2.0 & 70\%  & 10.95 & 1.41 \\
    \hline
    \end{tabular}
\end{table}

\section{An Application}

We compared our proposed smooth concordance metrics with the traditional C-Index using the Rotterdam breast cancer dataset, a well-known example for predictive survival modeling \citep{Foekens,Royston_Altman,survival-package}. In this analysis, our primary outcome was the time to breast cancer recurrence, and we built two different Cox proportional hazards models for comparison. In the initial model, we included predictors for age (years), menopausal status (post- versus pre-menopause), and hormonal therapy (yes or no). In the full model, we included all of the predictors from the initial model plus additional predictors for progesterone receptor levels (fmol/l) and the number of positive lymph nodes. 

\begin{figure}[h!]
    \centering
    \includegraphics[width=\textwidth]{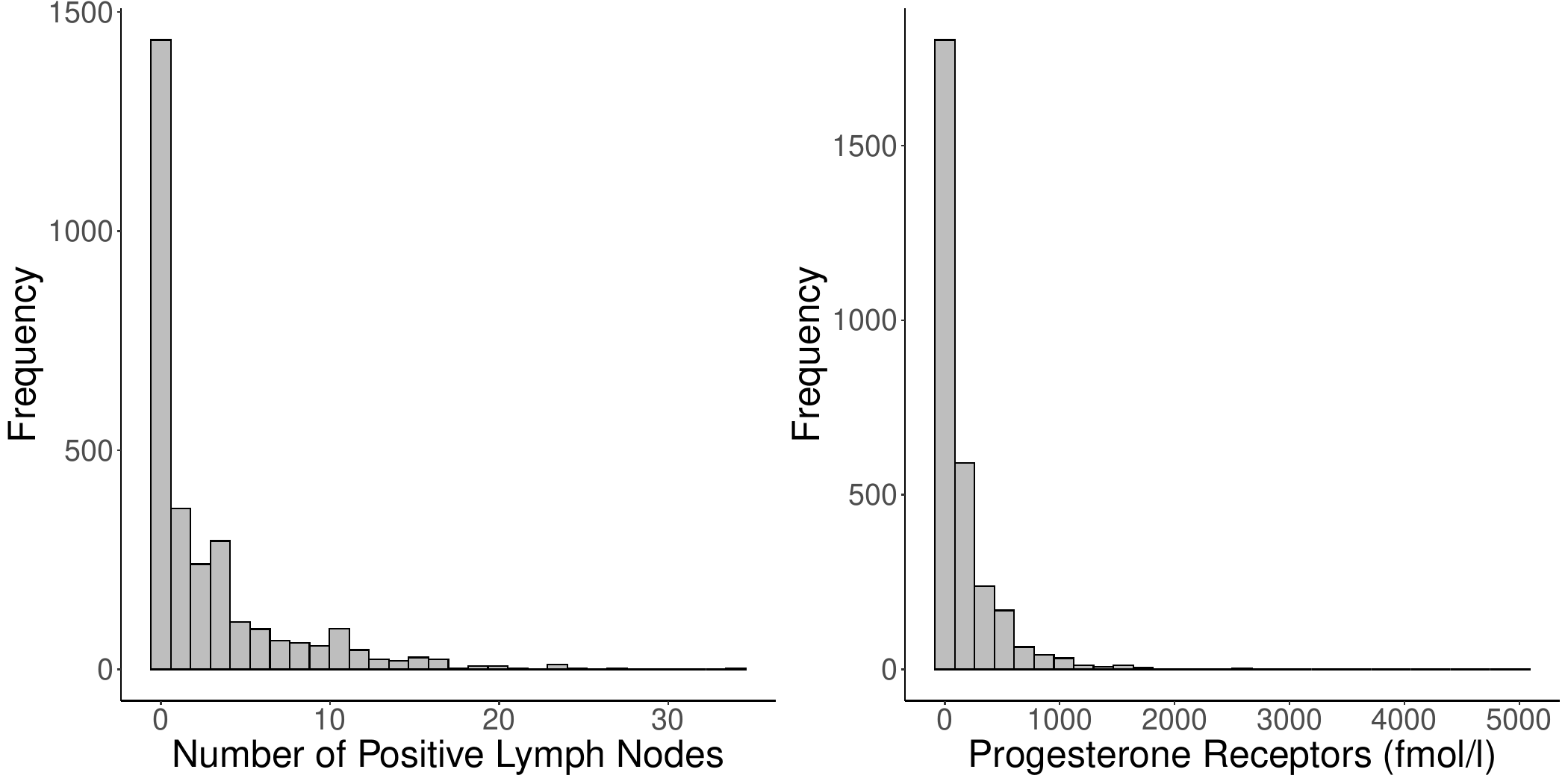}
    \caption{Histograms of the variables added into the full model for the Rotterdam breast cancer dataset.}
    \label{fig:hist}
\end{figure}

These additional variables that were added to the full model are known to be very strong predictors of breast cancer recurrence risk \citep{Schumacher,Foekens}. However, we observed that their distributions were highly skewed, with a tall peak at lower values and a long tail representing patients with highly extreme progesterone receptor levels or numbers of positive lymph nodes (Figure \ref{fig:hist}). Based on this information and the arguments in previous sections, we suspected that the traditional C-Index would understate the predictive power of the full model, whereas the smooth concordance metrics would appropriately reflect the useful information gained from the extreme tail of the risk score distribution. To assess this hypothesis, we estimated the traditional C-Index, $\exp(\nu)$, and the smooth sigmoid function for each model using the IPCW method to account for right-censoring. 

For the initial model, the estimated C-Index was 0.53 and the estimated odds ratio $\exp(\widehat{\nu})$ for a one standard deviation (SD) change in $R_i-R_j$ was 1.11, which corresponds to weak risk discrimination performance based on both metrics (because the values of the traditional C-Index and smooth concordance metrics that correspond to a useless risk score are 0.5 and 1.0 respectively). For the full model, the estimated C-Index was 0.64 and the estimated odds ratio $\exp(\widehat{\nu})$ was 1.88. While a C-Index of 0.64 is still considered to be low and likely would not support the use of a predictive survival model in practice \citep{DeMaris}, an odds ratio of 1.88 is very clinically meaningful, suggesting that every 1-SD increase in the risk score differences is associated with 88\% greater odds of one patient experiencing breast cancer recurrence before another. In other words, the $\exp(\widehat{\nu})$ estimate uncovers that the concordance probabilities increase rapidly with the differences in risk scores, and the progesterone receptor and lymph node variables are especially helpful for distinguishing large differences in risk that the initial model could not detect. 

\begin{figure}
    \includegraphics[width=\textwidth]{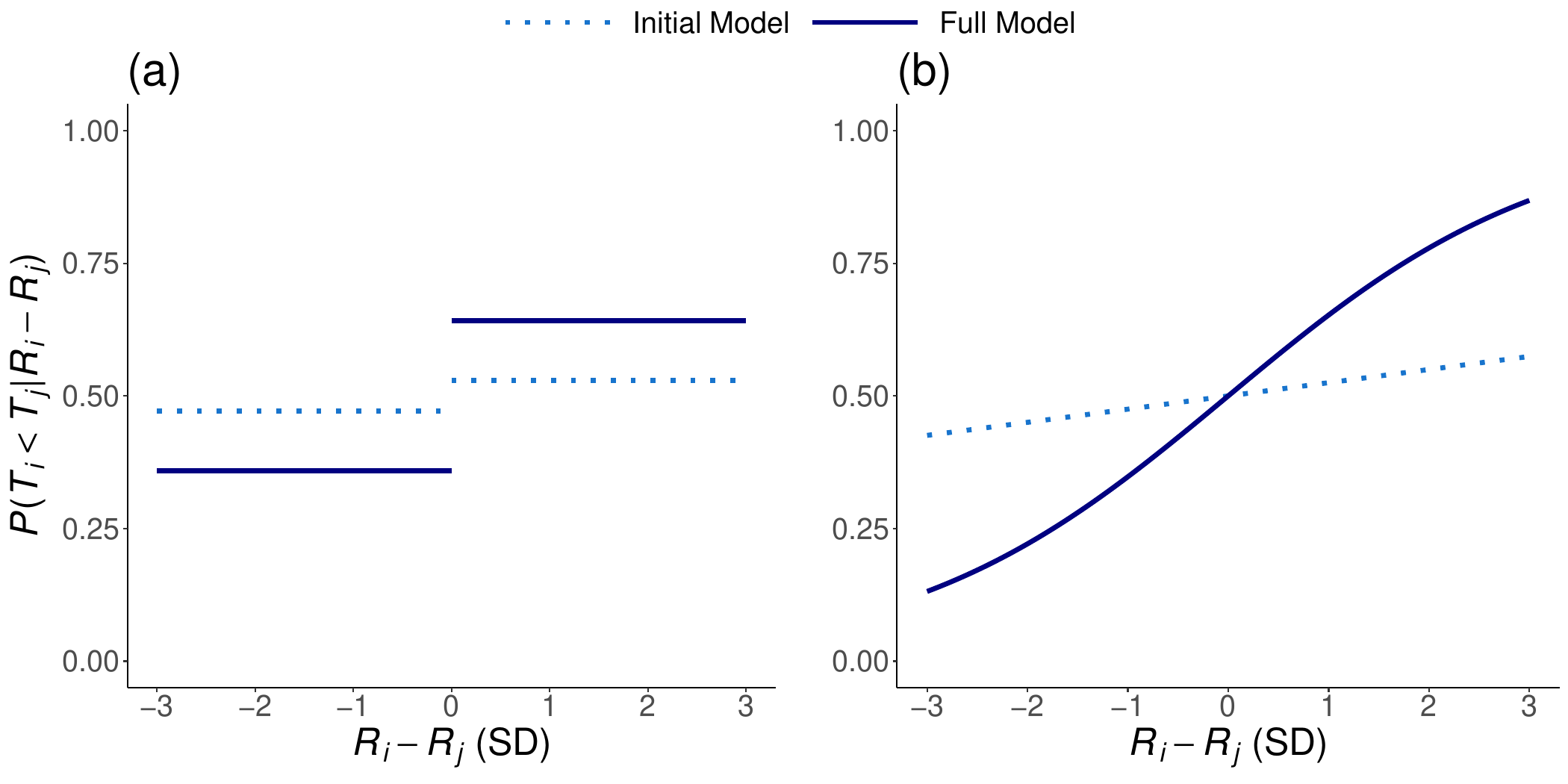}
    \caption{Probabilities of subject $i$ having an earlier underlying cancer recurrence time than subject $j$, based on the difference in risk scores $R_i-R_j$, for two different models fit to the Rotterdam breast cancer dataset. (a) probabilities based on the traditional concordance index and (b) probabilities based on the smooth concordance function. The initial model includes predictors for age (years), menopausal status (post- versus pre-menopause), and hormonal therapy (yes or no). The full model includes the predictors from the initial model plus additional predictors for progesterone receptor levels (fmol/l) and the number of positive lymph nodes. SD: Standard Deviation.}
    \label{fig:Rotterdam}
\end{figure}

Figure \ref{fig:Rotterdam} shows the estimated model performance as a function of the risk score differences, which helps further highlight the advantages of the smooth concordance metric under this example. The estimated sigmoid function in panel (b) shows that the full model has just minor improvements in concordance rates when the risk score difference $R_i-R_j$ is within one standard deviation of zero. However, for larger risk score differences (e.g., 2 or 3 SDs), the concordance rates are substantially higher in the full model compared to the initial model, reaching levels above 80\%. This reflects the fact that larger differences in progesterone receptor levels and the number of positive lymph nodes are very useful for discriminating underlying recurrence risk, and including these variables leads to major improvements in risk discrimination among these more extreme risk profiles. In contrast, the traditional C-Index marginalizes these probabilities over the distribution of risk score differences, which has the most density near zero. Therefore, it mainly describes the model performance among less clinically-meaningful comparisons and fails to detect the most important improvements that the full model offers. 

\section{Discussion}

In this paper, we have proposed a novel smooth concordance metric to assess risk discrimination performance for survival models across the entire range of possible differences in risk profiles. This metric addresses many of the well-known limitations of the traditional C-Index for right-censored survival data, offering increased sensitivity to the addition of strong predictors and better detection of useful discrimination performance for clinically meaningful risk differences. We have shown that smooth concordance assessments provide more information on risk discrimination performance by appropriately viewing concordance as a continuous function of risk score differences as opposed to a discrete function based on simple rank orderings. 

One advantage of the traditional C-Index is that it is a fully nonparametric statistic, whereas our proposed smooth concordance metric depends on semi-parametric models to link the concordance probability with the difference in risk scores. On the other hand, the sigmoid function in (\ref{eq:proposed}) is a flexible and canonical choice for this model, and the corresponding estimation approach under a weighted GEE allows for convenient selection of $h(\cdot)$ and corresponding model diagnostics. Furthermore, we suspect that these assumptions are often easier to satisfy in practice than those of \citet{Devlin} which depend on conditional survival functions, and we expect that even a misspecified model for (\ref{eq:proposed}) can be useful for uncovering the utility of a candidate risk score. 

With any model evaluation, it is essential that the selected performance metrics are relevant to the anticipated use of the model in practice. Our proposed smooth concordance metric is specifically relevant to risk discrimination scenarios where the focus is on distinguishing individuals who will experience the event of interest before other individuals. For example, medical professionals may be interested in risk discrimination to prioritize certain patients for additional treatments or scarce resources (e.g., organ transplants). In some contexts, the intended use of the survival model may be quite different, and other model evaluation statistics (e.g., calibration, net benefit, net reclassification) should be carefully considered \citep{Vickers2,Vickers}. 

Future extensions of this work may consider proposals of smooth concordance metrics for data structures beyond right-censored survival data, such as outcomes with competing risks, left-truncation, recurrent events, or interval censoring. As we have shown for the case of covariate-dependent right-censoring, the corresponding estimation strategy through GEE facilitates inverse probability weighting strategies to accommodate the complex data structures that arise in survival analysis. With these smooth concordance tools and future extensions, investigators will be able to more appropriately build and assess predictive survival models to ensure adequate performance prior to implementation in practice.

\bibliographystyle{apalike} 
\bibliography{main.bib}

\end{document}